\documentclass[twocolumn,showpacs,preprintnumbers,amsmath,amssymb]{revtex4}

\usepackage{graphicx}
\usepackage{dcolumn}
\usepackage{bm}

\begin{document}

\title{Floquet Spectrum and Transport Through an Irradiated Graphene Ribbon}

\author{Zhenghao Gu$^1$}
\author{H. A. Fertig$^1$}
\author{Daniel P. Arovas$^2$}
\author{Assa Auerbach$^3$}
\affiliation{1. Department of Physics, Indiana University, Bloomington, IN 47405}
\affiliation{2. Department of Physics, University of California at San Diego, La Jolla, CA 92093}
\affiliation{3. Department of Physics, Technion, 32000 Haifa, Israel}

\date{\today}

\begin{abstract}
Graphene subject to a spatially uniform, circularly-polarized electric field
supports a Floquet spectrum with properties akin to those of a topological
insulator, including non-vanishing Chern numbers associated with bulk bands
and current-carrying edge states.  Transport properties of this system however are
complicated by the non-equilibrium occupations of the Floquet states.  We address this
by considering transport in a two-terminal ribbon geometry for which the leads have
well-defined chemical potentials, with an irradiated central scattering region.
We demonstrate the presence of edge states, which for infinite
mass boundary conditions may be associated with only one of the two valleys.
At low frequencies, the bulk DC conductivity near zero energy is shown to be
dominated by a series of states
with very narrow anticrossings, leading to super-diffusive behavior.  For very
long ribbons, a ballistic regime emerges in which edge state transport dominates.

\end{abstract}

\pacs{72.80.Vp,73.23.-b,73.22.Pr} \maketitle

{\it Introduction and Key Results} --
The electronic properties of graphene are very unusual among two-dimensional
conducting systems, in large part because the low energy physics
is controlled by two Dirac points, which form the Fermi
surface of the system when undoped \cite{Castro_Neto_RMP,peres_2010,dassarma_2011}.
One of the very interesting possibilities for this system
is that, with spin-orbit coupling, it may represent the
simplest example of a topological insulator \cite{kane_2005,hasan_2010,qi_2010}.
Topological insulators are systems for which the bulk spectrum
is gapped, but which support robust, gapless edge states.  Unfortunately, spin-orbit
coupling in graphene appears to be too weak to allow observation of
this behavior with currently available samples.

Very recently, theoretical studies  have suggested that an analog of topological insulating
behavior can be {\it induced} in graphene by a time-dependent electric
potential \cite{oka_2009,lindner_2010,roslyak_2011,kitagawa_2010,kitagawa_2011,calvo_2011}.
The proposal entails exposing graphene to circularly-polarized electromagnetic
radiation of wavelength much larger than the physical sample size, such that only the electric field
has significant coupling to the electron degrees of freedom.  The periodic nature of the field necessitates
that the quantum states of the electrons are solutions of a Floquet problem,  characterized by a ``quasi-energy"
$\varepsilon_{\alpha}$ with allowed values in the interval $[-\omega/2,\omega/2]$,
where $\omega$ is the frequency of  the radiation \cite{razavy_2003}.  This same
physics allows one to induce topological-insulating properties in a variety of systems that are otherwise
only ``almost'' topological insulators \cite{lindner_2010}.  Because time-reversal symmetry is
explicitly broken in this system, it should support a Hall effect \cite{oka_2009} which, when measured
in an appropriate geometry, may be quantized \cite{kitagawa_2011}.

\begin{figure}
  \includegraphics[clip,width=6cm]{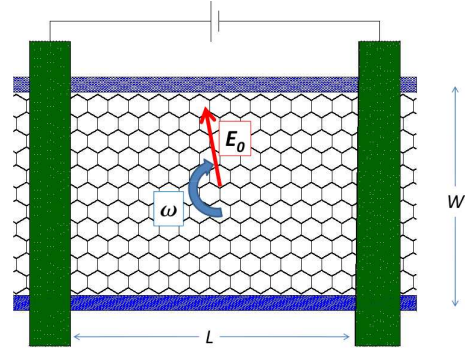}
  \caption{($Color$ $online$)
  Schematic diagram of device geometry.}
   \label{geometry}
\end{figure}

A key challenge one faces in determining transport properties of this system
is the assignment of electron occupations to the Floquet states.
In general, the quasienergies $\epsilon_\alpha$ cannot be simply inserted
as energies in a Fermi-Dirac distribution since they are limited to a finite
interval of real values, determined by the frequency.
In this study, we assume Fermi-Dirac distributions
only for the incoming waves far in the non-irradiated leads.
Thus we assume that electrons
are injected
and removed from the system via highly doped, ideal leads in which any possible effects
of an electric field have been screened out, and that the transport within the (finite) irradiated region
is quantum coherent.  Our geometry is a direct analog of one studied in Ref. \onlinecite{Tworzydlo_2006} for the time-independent case,
and is illustrated in Fig. \ref{geometry}.  Identical leads are taken to be made
of highly doped graphene.  This leads to a vanishing time-averaged current in
the absence of a DC bias, which avoids photovoltaic (charge pumping) effects.

\begin{figure}
  \includegraphics[clip,width=8cm]{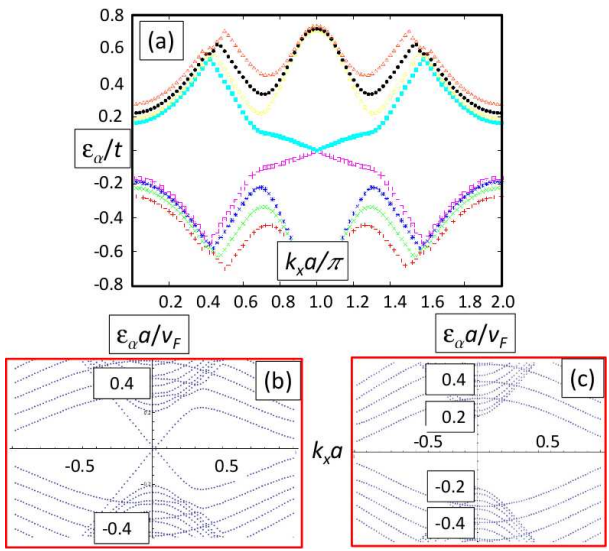}
  \caption{($Color$ $online$)
  (a) Tight-binding Floquet spectrum as a function of $k_x$ for a graphene ribbon oriented
  in the zigzag direction, subject to a circularly polarized electric field of
  magnitude $E_0$ with $eE_0a/\hbar\omega=0.5$, $\hbar\omega/t=3.0$, where $a$ is the Bravais
  lattice constant and $t$ the hopping parameter.
  (b,c) Spectra from Dirac equation with infinite mass boundary
  conditions, corresponding to two different valleys.  Only one valley
  carries edge states.  Frequency
  $\omega a/v_{\rm F}=5$, electric field amplitude $E_0/\hbar\omega=1$, ribbon width $W=45a$.}
   \label{bandstructure}
\end{figure}

For an infinite ribbon geometry, the momentum $k_x$ along the ribbon axis is a good
quantum number, and one may compute an effective band structure for the
Floquet eigenvalue $\varepsilon_{\alpha}$.  Fig. \ref{bandstructure} illustrates
this for several cases.  Fig. \ref{bandstructure}(a) displays the bands closest
to $\varepsilon=0$ for a relatively large frequency, $\hbar\omega=3t$
where $t$ is the tight-binding hopping parameter.  This
result is illustrative for its relative simplicity, but is only relevant
to very small system sizes for which the electric field may be approximated
as uniform throughout the sample \cite{kitagawa_2011}.  One may see two
sets of minima/maxima, corresponding to the two valleys, separated by a gap
that does not vanish even as the system width $W$ becomes very large.
This intrinsic gap in the Floquet spectrum is an analog of that which
opens in the presence of spin-orbit coupling for the static graphene
system \cite{kane_2005}.  In further analogy to this, a pair of edge
states traverses the gap and connects the two valleys, while (anti-)crossing
around $k_x=0$.

Figs. \ref{bandstructure}(b,c) illustrate corresponding results for Floquet spectra of
two individual Dirac points subject to the circulating electric field, corresponding to the two
different valleys, with infinite mass boundary conditions. In analogy with a topological insulator,
one sees that both valleys develop gaps which do not go away as $W$ becomes large,
but only one supports  edge states, so that the number of states at an edge is unaffected
(modulo 2) by the change of boundary condition, while the details of how these states
disperse are changed.  Since the system with its edge state may be well-described
qualitatively with just the single valley illustrated in Fig. \ref{bandstructure}(b),
we focus our attention on this case for the transport calculations.

\begin{figure}
  \includegraphics[clip,width=8cm]{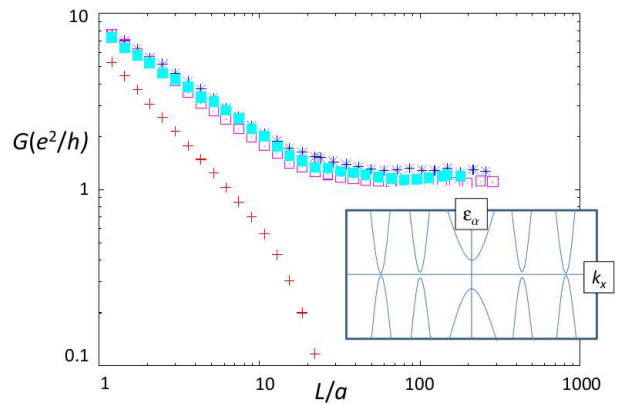}
  \caption{($Color$ $online$)
  Conductance vs. $L$ for several different bandwidths: $q_{n_{\rm max}}W= \pi n_{\rm max}$,
  with $n_{\rm max}=$ 10 (blue asterisks), 15 (lavender open squares), 25 (aqua closed squares).
  For these plots, $W=20a$, $\omega a/v_{\rm F}=5$, and number of time steps is 13.
  Note approximate power law behavior $G \sim L^{-b}$ with $b \approx 0.65$
  for $L < W$.  For comparison, results for $E_0=0$ displayed as red crosses,
  displaying $G \sim 1/L$ behavior for $L<W$.
  Inset: Illustration
  of $\varepsilon_\alpha$ vs. $k_x$ for infinite system with periodic boundary
  conditions, showing Floquet copies of spectrum and resulting avoided crossings
  for $k_x \ne 0$.}
   \label{data}
\end{figure}

Fig. \ref{data} illustrates typical results for the conductance of the
system as a function of the scattering region length, $L$.  At large
$L$ one can see the conductance level off to a constant value, indicating
the presence of edge states.  For sufficiently wide samples this ballistic behavior
will be robust against disorder, since states carrying current in
opposite directions reside on opposite sides of the sample.  The presence
of a finite current for large $L$ is in marked contrast to the behavior
in the absence of the radiation, for which the conductance vanishes \cite{Tworzydlo_2006}.
Surprisingly, the conductance exceeds the unirradiated conductance
for all values of $L$, in spite of the fact that a gap has opened in
the (Floquet) spectrum of the bulk system \cite{oka_2009}.  The explanation
of this lies in the fact that Floquet eigenvalues are restricted to a finite
interval, for example $-\omega/2<\varepsilon_{\alpha}<\omega/2$,
so that energy states outside this interval in the absence of the
periodic potential are folded into it.  For fixed $k_y$, this results
in a series of repeated
crossings at values of $v_{\rm F} k_x \approx m\omega$, where $m$
are non-zero integers.  In a system without edges
(i.e., if one considers periodic boundary conditions rather than a ribbon)
these become avoided crossings with very small gaps, as we explain
below, which transport current via evanescent states.  Remarkably,
transport through these states results in super-diffusive behavior,
$G \sim 1/L^b$, with $b<1$, as is apparent in Fig. \ref{data}.  This
non-analytic behavior reflects an explosive growth of the decay
length of evanescent states with large $|m|$, which has the form
\begin{equation}
\xi_m \approx \left( \frac{|m| \,\omega}{A_0} \right)^{\!2|m|}\frac{e^{-\eta\,|m|}}{|m| \, \omega},
\label{xi_m}
\end{equation}
where $A_0=E_0/\omega$ characterizes the electric field amplitude, assumed to be small, and
$\eta$ is a number of order unity.  As we argue below, the rapid growth
of $\xi_m$ with $m$ results in an anomalously large penetration of
the electrons into the irradiated region.

{\it Transmission Through Irradiated Region} -- The wavefunctions
for Dirac electrons subject to circularly polarized radiation
obey the time-dependent Schr{\"o}dinger equation
$[-i\partial_t+{\cal H}] \,\Psi\equiv H_{\rm F}\Psi=0$,
where the Hamiltonian of the system has the form
\begin{equation}
{\cal H} =
\left(
\begin{array}{cc}
0 & p_x-ip_y+A_x-iA_y  \\
p_x+ip_y+A_x+iA_y & 0
\end{array}
\right)
\label{hamiltonian}
\end{equation}
where ${\bf A}=A_0(\cos\omega t,\sin\omega t)$, and $p_{x,y}=-i\partial_{x,y}$.
(Note we have set the Fermi velocity $v_{\rm F}=1$ in this expression.)
Since ${\cal H}$ is periodic in time, the solutions will be Floquet states,
which have the form
$\Psi({\bf r},t)=e^{i\varepsilon_{\alpha}t} [\Phi_A({\bf r},t),\Phi_B({\bf r},t)]^{\textsf T}$, with
$\Phi_{\mu}({\bf r},t+T)=\Phi_{\mu}({\bf r},t)$ and $T=2\pi/\omega$.  Adopting
infinite mass boundary conditions leads to the conditions \cite{Tworzydlo_2006}
$\Phi_A(x,y=0,t)=\Phi_B(x,y=0,t)$ and $\Phi_A(x,y=W,t)=-\Phi_B(x,y=W,t)$.
Eigenstates of ${\cal H}$ which meet these boundary conditions are
\begin{eqnarray}
{\Phi}^{(n,s)}=\frac{1}{{\cal N}_s}\left \lbrace
\left(
\begin{array}{c}
sz^*-1 \\
1-sz
\end{array}
\right)
e^{iq_ny} +
\left(
\begin{array}{c}
1-sz \\
sz^*-1
\end{array}
\right)
e^{-iq_ny}
\right\rbrace \nonumber \\
\times e^{ik_xx-iA_yy},\quad\quad\quad\quad\quad\quad\quad\quad
\label{basis}
\end{eqnarray}
where $z=(q_x+iq_n)/q$, $q_x=k_x-A_x$, $q_n=(n+{1 \over 2})\pi/W$, $s = \pm 1$,
${\cal H} \Phi=sq\Phi$, and ${\cal N}_s$ is a normalization constant.
States in the leads of the system are generated by setting ${\bf A}=0$
in Eq. \ref{basis}.

Our strategy for finding the Floquet eigenvalues is to discretize time
and expand the Floquet operator $H_{\rm F}$ in instantaneous eigenstates of the Hamiltonian
${\cal H}$.  Writing $\Phi^{(n,t_i,s)}(t) \equiv \Phi^{(n,s)}(t) \, \delta_{t_i,t}$,
we may then write
$\langle n_1,s_1,t_1|{\cal H}(t) | n_2,s_2,t_2 \rangle =
E_{(n_1,s_1)}(t_1) \, \delta_{n_1,n_2}\delta_{s_1,s_2}\delta_{t_1,t_2}$.
The Floquet operator can them be written as a matrix of the form
\begin{eqnarray}
\langle n_1,s_1,t_1|H_{\rm F}|n_2,s_2,t_2 \rangle =
E_{(n_1,s_1)}(t_1) \, \delta_{t_1,t_2}\delta_{n_1,n_2}\delta_{s_1,s_2} \nonumber \\
-\frac{i}{2\Delta t}\Bigl[\langle n_1,s_1,t_1 | n_2,s_2,t_1-\Delta t \rangle \,
\delta_{t_2,,t_1-\Delta t}\quad\quad\quad \nonumber \\
- \langle n_1,s_1,t_1 | n_2,s_2,t_1+\Delta t \rangle \,
\delta_{t_2,t_1+\Delta t} \Bigr], \quad
\label{floquet_matrix}
\end{eqnarray}
whose eigenvalues are the allowed values of $\varepsilon_{\alpha}$
for an infinite ribbon.  Note that the states are implicitly functions of
$k_x$.  Diagonalization of Eq. \ref{floquet_matrix} generates results
such as those depicted in Fig. \ref{bandstructure}(b).  These
results were obtained for $0 \le n \le 30$ and 19 time slices.

Turning to the conductance, in the leads there are no microwaves, so that eigenvalues of
$H_{\rm F}$ with ${\bf A}=0$ have the form $\varepsilon=
\pm\sqrt{k_x^2+q_n^2}+\sin(m\omega \Delta t)/\Delta t \equiv \pm E_n(k_x)+ \varepsilon^t_m$.
This implicitly defines an equation for $k_x$, which depends on the integers
$n$ and $m$.  In the scattering region as well, for a given subband
of the Floquet spectrum, we need to know the values of $k_x$ that will
give some specified
Floquet eigenvalue
$\varepsilon_{\alpha}$.  This is equivalent to finding the values of $k_x$
where the bands illustrated in Fig. \ref{bandstructure}(b) cross some specified
horizontal line.  Note that if there is no such crossing for a given band
then the corresponding $k_x$ is actually complex, indicating an evanescent state.

In order to match the wavefunctions in the leads to the scattering
region we need to know these latter values of $k_x$, for a given
Floquet eigenvalue $\varepsilon_{\alpha}$.  To accomplish
this we multiply the eigenvalue equation by $\sigma_x$ to obtain
\begin{equation}
\big[(-i\partial_t-\varepsilon_{\alpha}) \sigma_x +i\sigma_zp_y\big]\vec{\psi}=k_x\vec{\psi}.
\label{kx_eq}
\end{equation}
This is a non-Hermitian matrix equation which we approximately solve in
a manner analogous to what we did for the original Floquet equation.
Eigenvectors give us the wavefunctions in the scattering region, and the
eigenvalues $k_x$ that enter into the plane wave part of the wavefunction
$e^{ik_xx}$.  A comparison of these solutions to $\varepsilon_{\alpha}(k_x)$
obtained by diagonalizing $H_{\rm F}$ reveals excellent agreement between
the two calculations.

We now have analytic formulas for the wavefunctions in the leads, and approximate
numerical solutions for them, represented in a finite basis of the states
$\Phi^{(n,s)}(y)\,e^{ik_x(n,m,s)x + imt}$, in the scattering region.  These need to
be matched at two junctions.  To accomplish this we match the wavefunctions on
discrete points in $y$, taking $y_j=(j+{1\over 2})W/(n_{\rm max}+1)$, where $n_{\rm max}+1$ is the
number of transverse states retained, and $j\in\{0,\ldots,n_{\rm max}\}$.
This defines a set of linear equations which we solve numerically, and
from which the matrix $T^{LR}_{qp}(E,E+\varepsilon^t_n)$,
representing the time-averaged transmission across the structure,
may be obtained. (Here $q,p$ represent transverse channels in the
left and right leads, respectively, and $E$ is the energy of an
impinging electron from the left.)
One may show that by matching on these particular points one enforces
current conservation in the solution; this is confirmed
numerically to 1 part in $10^3$.  The conductance is finally
given by \cite{datta_1997,comment_TR}
\begin{equation}
G=\frac{e^2}{h} \sum_{p,q,n} T^{LR}_{qp}(E_{\rm F},E_{\rm F}+\varepsilon^t_n)
\label{conductance}
\end{equation}
where $E_{\rm F}$ is the Fermi energy in the leads.

\textit{Evanescent Transmission in Irradiated System} -- A prominent result
from the calculations of the two-terminal conductance is an approximate
power law behavior $G \sim L^{-b}$ when $L < W$.  As $b$ is non-integral
this represents non-analytic behavior, and the fact that it emerges
well above the large $L$ value of $G$ suggests it is a result of
evanescent state transport.  This behavior turns out to be rather
natural when one accounts for higher order crossings at non-zero $k_x$
of the Floquet spectrum.  In a realistic situation, for example if the
impinging radiation is in the microwave regime, $\hbar\omega \sim 10^{-3}\,$eV,
one will have many such crossings since the graphene bandwidth ($\sim$ 1\,eV)
is relatively large.

To demonstrate the behavior, we consider a simpler
problem in which there is a half-space of irradiated graphene, and a half-space
of highly-doped, unirradiated graphene, joined across $x=0$, and we
adopt periodic boundary conditions in the transverse direction so
that there are no edge states.  For this situation $k_y$ is a
good quantum number.  States with zero energy are perfectly
backscattered in this case because there are no propagating states with zero
Floquet eigenvalue.  In the steady state situation there is a charge
density tail penetrating the irradiated graphene, which
has an approximate power law falloff with $x$.

To see this, we need to know how evanescent wavefunctions fall off with $x$
inside the irradiated graphene.  In the limit of vanishing microwave
amplitude $A_0$, the Floquet spectrum will have pairs of states crossing
$\varepsilon_{\alpha}=0$ at $k_x=\pm k_m^0=\pm \sqrt{(m\omega)^2-k_y^2}$.
The degeneracy is lifted for non-vanishing $A_0$, which we take to be
small compared to $\omega$.  The resulting anti-crossing will
occur at high order in perturbation theory, since the degeneracy occurs
for states with time dependence $\exp(\pm i m \omega t)$, whereas
the perturbation $V=A_0\,\big(\sigma^x \cos\omega t + \sigma^y \sin\omega t\big)$
connects states whose frequencies differ by a single unit of $\omega$.  Thus
the gap that opens at the crossing will be proportional to $(A_0/\omega)^{2m}$.
Since the inverse gap is essentially the localization length we wish to compute
this.  One approach is to use the resolvent operator $\hat{G}(z)=(z-\hat{H_{\rm F}})^{-1}$
with $\hat{H}_{\rm F}=\hat{H}_0+\hat{V}$.  Poles of the 2$\times$2 matrix
\begin{equation}
\tilde{G}=
\begin{pmatrix}
\langle \,m\,,-\,\big|\,{\hat G}\,\big|\,m\,,-\rangle & \langle \,m\,,-\,\big|\,{\hat G}\,\big|-m\,,+\rangle \\
\langle -m\,,+\,\big|\,{\hat G}\,\big|\,m\,,-\rangle & \langle -m\,,+\,\big|\,{\hat G}\,\big|-m\,,+\rangle
\end{pmatrix}
\label{resolvent}
\end{equation}
where $s=\pm$ indicates a particle-like (+) or hole-like (-) state, are the eigenvalues
of $H_{\rm F}$.  Expanding $\tilde{G}$ in
powers of $\hat{V}$ allows one to define a self-energy,
$\tilde{G}^{-1}=\tilde{G}_0^{-1}-\tilde{\Sigma}=z-\tilde{H}_0-\tilde{\Sigma}$, with $\tilde{H}_0=0$ for the
states of interest.  The diagonal components of $\Sigma$ simply shift the precise
location of the anticrossing on the $k_x$ axis and may be ignored.  To lowest
non-trivial order, the off-diagonal components are
$\Sigma_{\pm}=\langle m\,,-\big|\, \Sigma\,\big|-m,+\rangle=\langle -m\,,+\,\big|\,\Sigma\,\big|\,m,-\rangle^*$,
with
\begin{align}
&\Sigma_{\pm}= \left( \frac{(k_x+ik_y)A_0}{2k} \right)^{\!2m} \\
& \times  \!\!\!\! \prod_{n=-m+1}^{m-1}\left[
\langle n\,,+\,\big|\,{\hat G}\,^{(0)}\,\big| n\,,+ \rangle -
\langle n\,,-\,\big|\,{\hat G}\,^{(0)}\,\big|n\,,- \rangle \right]. \nonumber
\end{align}
Evaluating the matrix elements and setting $z=0$, one finds
$\left|\Sigma_{\pm}\right|=A_0^{2m} (m\omega)^{1-2m}\prod_n \Big[\left( {n\over m} \right)^2 -1\Big]^{-1}$.  In the
limit of large $m$, the product can be evaluated; noting that
$\xi_m^{-1} = 2|\Sigma_{\pm}|$, one arrives at the estimate in Eq. \ref{xi_m}
with $\eta=4(1-\ln 2)\approx 1.227$.

To see the connection with power law behavior, one needs to develop matching
conditions at the $x=0$ interface with these wavefunctions.  This is a tedious
but in principle straightforward exercise \cite{herb_unpub}, yielding the
result that the contribution to the density from large $n$ has the
time-averaged form
$$\rho(x) \sim \sum_n \left(\frac{A_0}{\omega \tau}\right)^{\!2n} e^{-x/\xi_n},$$
where $\tau$ is of order unity, and depends on $k_y$.
The sum may be estimated by assuming it is dominated by a single term
at large $n$ when $x$ is large; maximization yields
$$
n_{\rm max} = \frac{\ln(\omega x)}{y+2\ell_n}+C
$$
where $C$ is independent of $L$, $\ell_n$ is a correction of order $\ln\!\big[\ln(\omega x)\big]$,
and $y=-2\ln(A_0/\omega)-\eta$.  Using this term to estimate the
sum yields the result $\rho(x) \sim (\omega x)^{2\ln(a/\omega\tau)/(y+2\ell_n)}$.
Since this is a weak function of $k_y$ (through $\tau$), we see that summing over
the transverse modes should result in an approximate power law density tail.
Such a fall-off is expected to lead to similar behavior in the transmission.
It is interesting to note that the
actual value of the exponent is relatively insensitive to $\omega$ and $A_0$,
since these enter only through logs; this is consistent with our results, for
which the observed power tends to remain in the interval $0.6<b<0.75$ over a variety
of choices for $\omega$ and $A_0$.

In summary, two-terminal transport through an undoped graphene ribbon subject to a circularly
rotating electric field has a conductance that reveals the unusual nature of the
Floquet spectrum of this system.  Evanescent transport through relatively short
ribbons is super-diffusive due to a series of near crossings with very small gaps.
At larger ribbon lengths, transport becomes ballistic, revealing the presence of
edge states which are a hallmark of the topological nature of the spectrum.

{\it Acknowledgements} -- This work was supported by the NSF through
Grant Nos. DMR-1005035 (HAF) and DMR-1007028 (DPA), and the US-Israel
Binational Science Foundation (AA and DPA).
We acknowledge the hospitality of the Aspen Center for Physics
where this work was initiated.  The authors thank Erez Berg,
Luis Brey, Fernando de Juan, Netanel Lindner and Gil Refael for helpful
discussions.


\begin{thebibliography}{17}
\bibitem[{com()}]{comment_TR}
\bibinfo{note}{In a general structure subject to a time-dependent potential
  there may be current even in the absence of DC bias across the system. One
  may show for that this is not the case the case for this structure, so that
  for small biases $V$ the time-averaged current is $GV$.}
\end{thebibliography}

\end{document}